\documentclass{article}

\usepackage[pdftex]{graphicx}
 
 \usepackage{spconf,amsmath,subcaption, makecell, array}
 \usepackage[nocompress]{cite}
 \usepackage{nicematrix}
\usepackage{enumitem}
\usepackage{algorithmic}
\usepackage{algorithm}

\newcolumntype{L}[1]{>{\raggedright\let\newline\\\arraybackslash\hspace{0pt}}m{#1}}
\newcolumntype{C}[1]{>{\centering\let\newline\\\arraybackslash\hspace{0pt}}m{#1}}
\newcolumntype{R}[1]{>{\raggedleft\let\newline\\\arraybackslash\hspace{0pt}}m{#1}}
\makeatletter
\newcommand{\ssymbol}[1]{^{\@fnsymbol{#1}}}
\makeatother

\ninept

\title{Efficient Robust Adaptive Beamforming Based on Spatial Sampling with Virtual Sensors }
\name{Saeed Mohammadzadeh \qquad  Rodrigo C. de Lamare }

\address{ University of York, UK and 
CETUC/PUC-Rio, Rio de Janeiro, Brazil}
\begin{document}
%
\maketitle
\begin{abstract}
Robust adaptive beamforming (RAB) based on interference-plus-noise covariance (IPNC) matrix reconstruction can experience serious performance degradation in the presence of look direction and array geometry mismatches, particularly when the input signal-to-noise ratio (SNR) is large. In this work, we present a RAB technique to address covariance matrix reconstruction problems. The proposed method involves IPNC matrix reconstruction using a low-complexity spatial sampling process (LCSSP) and employs a virtual received array vector. In particular, we devise a power spectrum sampling strategy based on a projection matrix computed in a higher dimension. A key feature of the proposed LCSSP technique is to avoid reconstruction of the IPNC matrix by integrating over the angular sector of the interference-plus-noise region. Simulation results are shown and discussed to  verify the effectiveness of the proposed LCSSP method against existing approaches. 
\end{abstract}
\section{Introduction}
\label{sec:intro}
Many adaptive beamforming methods have been applied in wireless communications, sonar, and radar due to their superior interference mitigation capability  \cite{van2004detection}. However, under non-ideal conditions such as finite data samples and mismatches between the presumed and true steering vector (SV) the performance of adaptive beamformers degrades substantially.  Several RAB techniques have been proposed to enhance robustness against the aforementioned mismatches, such as the linearly constrained minimum variance (LCMV) beamformer \cite{monzingo2004introduction}, diagonal loading (DL) \cite{kukrer2014generalised,l1stap,jiodoa,alrd,rprec,rpreccf}, the eigenspace-based beamformer  \cite{huang2012modified,rrstap}, the worst case-based technique \cite{vorobyov2003robust,rcb}, the probabilistically constrained approach in \cite{vorobyov2008relationship,rddrls} and the modified robust Capon beamformer in \cite{mohammadzadeh2018modified}. Hence, the development of low-complexity RAB approaches has been a very active research topic in recent years. Nevertheless, a major cause of performance degradation in adaptive beamforming is the presence of the desired signal component in the training data, especially at high SNR.

To address this issue, many works tried to remove the signal-of-interest (SOI) components by reconstruction of the interference-plus-noise covariance (IPNC) matrix instead of using the sample covariance matrix (SCM). In \cite{gu2012robust}, the IPNC matrix is reconstructed by integrating the nominal SV and the corresponding Capon spectrum over the entire angular sector except the region near the SOI. Several categories of IPNC matrix-based beamformers were then proposed, such as the beamformer in \cite{chen2015robust}, which relies on a correlation coefficient method, the computationally efficient algorithms via low complexity reconstruction in \cite{ruan2014robust,ruan2016,lrcc}, subspace-based algorithms \cite{jio,jidf,sjidf,mohammadzadeh2018adaptive}, an approach based on spatial power spectrum sampling (SPSS) \cite{zhang2016interference}, and the algorithm in \cite{chen2018adaptive} which constructs an IPNC matrix directly from the signal-interference subspace. The robust beamformer in \cite{gu2019adaptive} utilizes the orthogonal subspace (OS) to eliminate the component of the SOI from the angle-related bases while in \cite{Saeed2020} a robust beamformer is proposed based on the principle of maximum entropy power spectrum (MEPS) to reconstruct the IPNC and the desired signal covariance matrices.

In this paper, we develop an effective RAB approach that achieves nearly optimal performance by addressing the inaccurate covariance matrix construction problems with less computations than other approaches in the literature. The essence of the idea is based on IPNC matrix reconstruction using a low-complexity spatial sampling process (LCSSP) and employing virtual sensors. The power spectrum sampling is realized by a proposed projection matrix in a higher dimension. In contrast to previously reported works with IPNC construction, we avoid the reconstruction and estimation of the IPNC matrix by integrating over the angular sector of the interference-plus-noise region. Simulation results are presented to verify the effectiveness of the proposed method while requiring less computational complexity.

This paper is structured as follows. Section 2 introduces the system model and states the problem. Section 3 presents the proposed LCSSP method. Section 4 depicts and discusses the simulation results, whereas Section 5 draws the conclusions.

\section{Problem Background}
\label{sec:methodology}
Consider a linear antenna array of $M$ sensors with interelement spacing $d$. The data received at the $t^{th}$  snapshot depicted as $\mathbf{x}(t)= \mathbf{x}_\mathrm{s}(t)+\mathbf{x}_\mathrm{i}(t)+\mathbf{x}_\mathrm{n}(t) $ which is modeled by  
\begin{align}\label{Received Data Vector}
\mathbf{x}(t)= s(t)\mathbf{a}(\theta_\mathrm{s})+\sum_{p=1}^P i_\mathrm{p}(t) \mathbf{a}(\theta_\mathrm{p})+\mathbf{x}_\mathrm{n}(t),
\end{align}
where $P$ is the number of interfering signals. $s(t)$, $i_\mathrm{p}(t)$ and $\mathbf{x}_\mathrm{n}(t)$ denote the desired signal, interference signal waveform and noise components, respectively. Assume that the desired signal, interference, and noise are statistically independent from each other. $\theta_\mathrm{s}$, $\theta_\mathrm{p}$ denotes the direction of the desired signal and the $p$th interference, respectively. The vector $\mathbf{a}(\cdot)$ is the corresponding SV, which has the form $\mathbf{a}(\theta)=\frac{1}{\sqrt M} \Big[1, \ e^{j
2\pi\bar{d} \sin\theta},  \ \cdots, \ e^{j
2\pi (M-1)\bar{d}\sin\theta}\Big]^\mathrm{T}$,
where $\bar{d}=d/\lambda$=1/2, $\lambda$ is the wavelength, and $ (\cdot)^\mathrm{T} $ denotes the transpose. Assuming that the SV $ \mathbf{a}(\theta_\mathrm{s}) $ is known, then for a given beamformer weight vector $ \mathbf{w} $, the beamformer performance is measured by the output signal-to-interference-plus-noise ratio (SINR) as follows
\begin{align}\label{SINR}
\mathrm{SINR}=\sigma^{2}_\mathrm{s} |\mathbf{w}^\mathrm{H} \mathbf{a}(\theta_\mathrm{s})|^2 \big{/} \mathbf{w}^\mathrm{H} \mathbf{R}_\mathrm{i+n}\mathbf{w},
\end{align}
where $ \sigma^{2}_\mathrm{s} $ is the desired signal power, $ \mathbf{R}_\mathrm{i+n}=\mathbf{R}_\mathrm{i}+\mathbf{R}_\mathrm{n} $ is the IPNC matrix and $ (\cdot)^\mathrm{H} $ stands for Hermitian transpose. Assuming that the interfering signals are independent, the covariance matrix of the received signal vector is given by
\begin{align}\label{Theoretical R}
    \mathbf{R}=  \sigma^{2}_\mathrm{s}\mathbf{a}(\theta_\mathrm{s})\mathbf{a}^\mathrm{H}(\theta_\mathrm{s})+\sum_{p=1}^P \sigma^{2}_\mathrm{p}\mathbf{a}(\theta_\mathrm{p})\mathbf{a}^\mathrm{H}(\theta_\mathrm{p})+\sigma^{2}_\mathrm{n}\mathbf{I}   
\end{align}
where $\sigma^2_\mathrm{n}$, $\sigma^2_\mathrm{p}$ and $\mathbf{I}$ represent the power of the white Gaussian noise, of each interference component and the identity matrix, respectively. Assuming that the SV $\mathbf{a}(\theta_\mathrm{s})$ is known precisely, the problem of maximizing the SINR in (\ref{SINR}) can be cast as the following optimization
problem:
\begin{align}\label{MVDR}
\underset{{\mathbf{w}}}{\operatorname{min}}\ \mathbf{w}^\mathrm{H} \mathbf{R}_\mathrm{i+n} \ \mathbf{w}\ \hspace{.4cm} \mathbf{s.t.} \hspace{.4cm} \mathbf{w}^\mathrm{H} \mathbf{a}(\theta_\mathrm{s})=1.
\end{align} 
The solution to (\ref{MVDR}) yields the optimal beamformer given by
\begin{align}\label{optimal wegight vector}
\mathbf{w}_{\mathrm{opt}}=\mathbf{R}_\mathrm{i+n}^{-1} \mathbf{a}(\theta_\mathrm{s}) \big{/} \mathbf{a}^\mathrm{H}(\theta_\mathrm{s}) \mathbf{R}_\mathrm{i+n}^{-1}\mathbf{a}(\theta_\mathrm{s}).
\end{align}
However, in practice the exact IPNC matrix, $ \mathbf{R}_\mathbf{i+n} $ and array covariance matrix, $ \mathbf{R} $ are unavailable even in signal-free applications, thus they are replaced by the SCM, $ \hat{\mathbf{R}}=(1/K)\sum_{t=1}^{K} \mathbf{x}(t)\mathbf{x}^\mathrm{H}(t) $, where  $K$ is the number of snapshots.  
 
\section{Proposed LCSSP Algorithm}
\indent When the incident signal is narrowband, the signal varies slowly with time (assuming that the carrier has been removed). In the noise-free case, a single snapshot is adequate as it contains all available information. A snapshot of a narrowband signal $\beta(t)$ arriving from direction $\phi$ \cite{mohammadzadeh2019robust} may be expressed as 
\begin{align}
    \mathbf{x}(t)= \beta(t) \mathbf{a}(\phi).
\end{align}
The vector representation of the array output model as in \eqref{Received Data Vector} plays a very crucial role in the development of high-resolution methods for direction of arrival (DoA) estimation. To steer an array to the desired direction, $\phi_0$, we form an inner product of the SV and the array snapshot \cite{naidu2009sensor}
\begin{align}
    \mathbf{a}^\mathrm{H}(\phi_0)\mathbf{x}(t)= \beta(t)\mathbf{a}^\mathrm{H}(\phi_0)\mathbf{a}(\phi).
\end{align}
Therefore, the response of the steered array is expressed as an inner product of the SV and the direction vector as
\begin{align}  \label{f function}
    g(\phi;\phi_0)=&\mathbf{a}^\mathrm{H}(\phi_0)\mathbf{a}(\phi)=\frac{1}{M}\sum_{m=0}^{M-
    1}e^{jm\pi [\sin(\phi)-\sin(\phi_0)]}, \qquad \raisetag{23pt}
\end{align}
where $\phi_0, \phi \in [-\pi/2, \pi/2]$  are the desired direction to which the array is steered and the DoA of a wavefront, respectively. The function $g(\phi;\phi_0)$ is called the selection function or indication function of the SV \cite{zhang2016interference}. Letting $z=[\sin(\phi)-\sin(\phi_0)]M/2
\in [(-1-\sin(\phi_0))M/2, \ (1-\sin(\phi_0))M/2)] $, equation \eqref{f function} can be rewritten as 
\begin{align}\label{F(z)}
    g(z)= \dfrac{1}{M}\sum_{m=0}^{M-1}e^{j(2 \pi/M)m z}=\dfrac{\sin (\pi z)}{M\sin(\pi z/M)} e^{j\frac{M-1}{M} \pi z}. \qquad \raisetag{23pt}
\end{align}
Note that $\lim_{z\rightarrow 0} g(z) = 1$, and if $\phi_0\in(-\pi/2, \pi/2)$, the denominator of \eqref{F(z)} will be zero only for $z=0$ ($\phi = \phi_0$), so $g(z)$ has $M-1$ zeros in the interval $[(-1-\sin(\phi_0))M/2, (1-\sin(\phi_0))M/2)$.  
We denote these $M-1$ zeros of $g(z)$ and $g(\phi;\phi_0)$ as  $z_m=[\sin(\phi_m)-\sin(\phi_0)]M/2$ with $\phi_m=\arcsin (2z_m/M+\sin(\phi_0))$, for $m=1,\cdots, M-1 $ respectively. The set $\{\mathbf{a}(\phi_m)\}_{m=0}^{M-1}$ is linearly independent (to see this, note that $\mathbf{A} = \begin{bmatrix}\mathbf{a}(\phi_0) &\dots & \mathbf{a}(\phi_{M-1}) \end{bmatrix}$ is a Vandermonde matrix with different columns if $\bar{d}\le 1/2$). This ensures that the SVs of $(\phi_0,\lbrace \phi_m \rbrace _{m=1}^{M-1})$ form an orthonormal basis that spans the $M$-dimensional complex space.\\
We intend to construct a projection matrix orthogonal to the signal subspace that retains as much as possible the interference-plus-noise (see \eqref{C_L} below). To this end, it is convenient to extend the array with a number of virtual sensors, as explained next.
In Fig.~1, the design of the extended array is shown with $L-M$ virtual sensors.
\begin{figure}[!] \label{Hypothetical array}
	\centering
	\includegraphics[width=.36\textwidth]{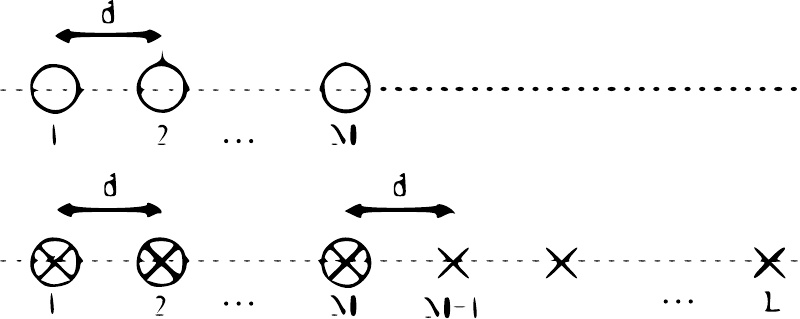}
\vspace{-0.75em}
	\caption{Virtual and physical array configuration}
	\label{Beampattern}
\end{figure}

Let us define an extended received vector as 
\begin{align}
      \mathbf{x}_\mathrm{L}(t)=
      \begin{bmatrix}
      \mathbf{x}(t)\\
      \mathbf{x}_\mathrm{e}(t)
      \end{bmatrix},
\end{align}
where $\mathbf{x}(t)=[x_0(t), x_1(t), \cdots, x_{M-1}(t)]^\mathrm{T}$ is the actual received vector of the array (of dimension $M$), and the vector received by the virtual part of the extended array (of dimension $L-M$) is depicted as $\mathbf{x}_\mathrm{e}(t)=[x_{M}(t), x_{M+1}(t), \cdots, x_L(t)]^\mathrm{T}$. Moreover, for the new dimension ($L$) the corresponding SV of the desired signal $\mathbf{a}_\mathrm{L}(\theta_\mathrm{s})$ and the interferences $\mathbf{a}_\mathrm{L}(\theta_\mathrm{p})$ are depicted as $\mathbf{a}_\mathrm{L}(\theta)=\frac{1}{\sqrt L} \Big[1, \ e^{j
2\pi\bar{d} \sin\theta},  \cdots,  e^{j
2\pi (L-1)\bar{d}\sin\theta}\Big]^\mathrm{T}$. Thus, the covariance matrix of the extended array’s received vector is
\begin{align}\label{R_L}
  \mathbf{R}_\mathrm{L}= &E\{\mathbf{x}_\mathrm{L}(t) \ \mathbf{x}^\mathrm{H}_\mathrm{L}(t) \} \nonumber \\ \hfill
      = &E\bigg\{
      \begin{bmatrix}
      \mathbf{x}(t)\\
      \mathbf{x}_\mathrm{e}(t)
      \end{bmatrix} 
      \begin{bmatrix}
      \mathbf{x}^\mathrm{H}(t)& \mathbf{x}^\mathrm{H}_\mathrm{e}(t)
      \end{bmatrix}
      \bigg\} =
      \begin{bmatrix}
      \mathbf{R}& \mathbf{R}_1\\
      \mathbf{R}^\mathrm{H}_1& \mathbf{R}_2
      \end{bmatrix},
\end{align}
where $\mathbf{R}$ is the covariance matrix of the original array.

The set of orthogonal SVs with dimension $L$ corresponding to the angles $(\phi_0,\lbrace \phi_{\ell} \rbrace _{\ell=1}^{L-1})$ spans the $L$-dimensional complex space.
In the following, we show that the IPNC matrix is accurately estimated without resorting to the power spectrum of the interference SVs directly. Instead, we define a projection matrix $\mathbf{C}_\mathrm{L}$ onto an approximation to the orthogonal space of the SV of the SOI, taking advantage of the higher dimensions to better approximate the interference-plus-noise space. Assuming $\phi_0\approx\theta_\mathrm{s}$, we use the set of all angles  outside the SOI region  $\Phi= \lbrace \phi_1, \phi_2, \cdots, \phi_{\ell}, \cdots, \phi_{L-1} \rbrace$ to define the orthogonal projection basis. The angles in $\Phi$ can then be used to generate the projection matrix as 
\begin{align} \label{C_L}
    \mathbf{C}_\mathrm{L}= \sum_{\phi_{\ell} \in \Phi} \mathbf{a}(\phi_{\ell}) \mathbf{a}^\mathrm{H}(\phi_{\ell}),
\end{align}
where $\mathbf{C}_\mathrm{L}$ is the orthogonal projection matrix onto the span of  $\mathbf{a}(\phi_1), \mathbf{a}(\phi_1), \cdots, \mathbf{a}(\phi_{L-1})$. Note that if we choose $\phi_0\approx\theta_\mathrm{s}$, $\mathbf{C}_\mathrm{L}\mathbf{a}_\mathrm{L}(\theta_\mathrm{s})\approx\mathbf{0}$ and if $L$ is sufficiently large, $\mathbf{C}_\mathrm{L}\mathbf{a}_\mathrm{L}(\theta_{p})\approx\mathbf{a}_\mathrm{L}(\theta_{p})$, $\mathbf{C}_\mathrm{L}\mathbf{x}_n\approx\mathbf{x}_n$, as long as the directions $\theta_{p}$ of the interferers are not too close to the direction of the desired signal $\theta_\mathrm{s}$.
To compute the parameter $L$ in the corresponding interfering signals, the quadratic error can be minimized in these directions as follows
\begin{align}
    \epsilon=&\underset{{\mathbf{C}_\mathrm{L}}}{\operatorname{min}} \sum _{p=1}^P \parallel \mathbf{C}_\mathrm{L}\mathbf{a}_\mathrm{L}(\theta_p)- \mathbf{a}_\mathrm{L}(\theta_p) \parallel ^2 \nonumber \\ =&\parallel \mathbf{C}_\mathrm{L}\mathbf{B}(\theta)-\mathbf{B}(\theta) \parallel ^2_{F},
\end{align}
where $\mathbf{B}(\theta)=[\mathbf{a}_\mathrm{L}(\theta_1), \cdots, \mathbf{a}_\mathrm{L}(\theta_P)
]$ and  $\|\cdot\|_F$ denotes the Frobenius norm. Defining the normalized error as 
\begin{align}\label{eq:error}
    \epsilon_n(\mathbf{C}_\mathrm{L})=\dfrac{ \operatorname{min}\parallel \mathbf{C}_\mathrm{L}\mathbf{B}(\theta)-\mathbf{B}(\theta) \parallel_F }{\parallel  \mathbf{B}(\theta) \parallel_F}.
\end{align}
the precision can be guaranteed if the error is limited to a small value $\delta$. The best choice for $L$ can be found by computing the error \eqref{eq:error}, starting from $L=M$ and increasing $L$ until the error is smaller than a prescribed threshold  $\delta $.

Let us define 
\begin{align}\label{eq:CLapprox}
    \mathbf{u}=\mathbf{C}_\mathrm{L} \mathbf{x}_\mathrm{L}&=\dfrac{1}{L}\Big(\sum _{\ell=1}^{L-1} \mathbf{a}(\phi_\ell)\mathbf{a}^\mathrm{H}(\phi_\ell)\Big)\Big(s(t)\mathbf{a}_\mathrm{L}(\theta_\mathrm{s})+ \nonumber  \sum_{p=1}^P  i_\mathrm{p}(t) \\& \cdot \mathbf{a}_\mathrm{L}(\theta_{p})+\mathbf{x}_\mathrm{n}\Big) 
    \approx\sum_{p=1}^P i_\mathrm{p}(t)\mathbf{a}_\mathrm{L}(\theta_{p})+\mathbf{x}_\mathrm{n}.
\end{align}

Using these approximations,  post-multiplying the above equation by $\mathbf{u}^\mathrm{H}$ and taking the expected value of $\mathbf{u}\mathbf{u}^\mathrm{H}$, assuming that the interferers $i_\mathrm{p}(t)$ are uncorrelated with each other, the signal $s(t)$ and with the noise $\mathbf{x}_n$, we have
\begin{align}
    E\{\mathbf{u}\mathbf{u}^\mathrm{H}\}&=E\{\mathbf{C}_\mathrm{L} \mathbf{x}_\mathrm{L} \mathbf{x}_\mathrm{L}^\mathrm{H} \mathbf{C}_\mathrm{L}^\mathrm{H}\} \nonumber\\
    &\approx E\Big{\lbrace}  \big (\sum_{p=1}^P i_\mathrm{p}(t)\mathbf{a}_\mathrm{L}(\theta_{p})+\mathbf{x}_\mathrm{n} \big) \big(\sum_{j=1}^P i_j(t)\mathbf{a}_\mathrm{L}(\theta_j)+\mathbf{x}_\mathrm{n}\big)^\mathrm{H} \Big{\rbrace}  \nonumber\\
  &=\sum_{p}\sigma_\mathrm{p}^2\mathbf{a}_\mathrm{L}(\theta_{p})\mathbf{a}_\mathrm{L}^\mathrm{H}(\theta_{p})+\sigma_\mathrm{n}^2\mathbf{I}.\raisetag{20pt}
\end{align}
We can also write
\begin{equation}\label{Final R_i+n}
\begin{split}
E\{\mathbf{u}\mathbf{u}^\mathrm{H}\}&=\mathbf{C}_\mathrm{L}  \mathbf{R}_\mathrm{i}  \mathbf{C}_\mathrm{L}^\mathrm{H}+ \mathbf{C}_\mathrm{L}  \mathbf{R}_\mathrm{n}  \mathbf{C}_\mathrm{L}^\mathrm{H} \\
     &=\mathbf{C}_\mathrm{L}  \mathbf{R}_\mathrm{i+n}^\mathrm{(L)}  \mathbf{C}_\mathrm{L}^\mathrm{H}
     =\begin{bmatrix}
      \hat{\mathbf{R}}_\mathrm{i+n}& \mathbf{X}_1\\
      \mathbf{X}_1^\mathrm{H}& \mathbf{X_2}
      \end{bmatrix}.
\end{split}\end{equation}
The first $M$ entries in $\mathbf{x}_\mathrm{L}$ correspond to the physical sensors, which originate the estimate of the $M\times M$ IPNC matrix, $\hat{\mathbf{R}}_{\text{i+n}}$, in \eqref{Final R_i+n}. The matrices $\mathbf{X}_1$ and $\mathbf{X}_2$ correspond to the virtual part of the array. The approximation in \eqref{eq:CLapprox} is effective when $L>M$, resulting in an improved estimate of the IPNC matrix $\hat{\mathbf{R}}_\mathrm{i+n}$.   
The robust beamformer is computed by 
\begin{align}\label{proposed wegight vector}
\mathbf{w}_{\mathrm{prop}}=\dfrac{\hat{\mathbf{R}}_\mathrm{i+n}^{-1} \mathbf{a}(\bar{\theta}_\mathrm{s})}{\mathbf{a}(\bar{\theta}_\mathrm{s})^\mathrm{H} \hat{\mathbf{R}}_\mathrm{i+n}^{-1}\mathbf{a}(\bar{\theta}_\mathrm{s})}.
\end{align}
\begin{algorithm} [t]
	\caption{Proposed LCSSP Robust Adaptive Beamforming }\label{sobelcode}
	1: \textbf{Initialization:}\:\textit{L}=\textit{M}, $\bar{\theta}_s$, $\phi_0$, $\delta$,  $ \theta_1, \cdots, \theta_p$. \\
	2: \textbf{Input} $\mathbf{a}_\mathrm{L}(\theta)$,  $\mathbf{B}$  \\
	3: \textbf{For} $\textit{L}=\textit{M}:1:\cdots: \text{Do}$\\
	4: Compute Array received data vector $\lbrace \textbf{{x}}_\ell(t) \rbrace_{\ell=1}^L $,\\
	5: Compute  ${\hspace{1em}}\hat{\textbf{R}}_\mathrm{L}=(1/K)\sum_{t=1}^{K} \textbf{{x}}_\mathrm{L}(t)\textbf{{x}}_\mathrm{L}^H(t)$;   \\
	6: Define interval as $\text{Int}=[(-1-\sin(\phi_0))L/2, \  (1-\sin(\phi_0))L/2)]$\\
	7: Define $z=\text{Int}(1):\text{Int}(2)$\\
	8: \qquad   $\textbf{For}$ $\ell=1:\text{length}(z)$\\
	9: \qquad \qquad $\phi(\ell)=\arcsin (2z(\ell)/L+\sin(\phi_0))$\\
	10: \qquad \qquad \qquad $\textbf{If}$ \qquad $\phi(\ell) \in \bar{\Theta}$ \qquad $\textbf{then}$ \\
	11:   \qquad   \qquad \qquad $\mathbf{C}_\mathrm{L}= \sum_{\phi(\ell)}  \mathbf{a}_\mathrm{L}(\phi(\ell)) \ \mathbf{a}_\mathrm{L}^\mathrm{H}(\phi(\ell))$  \\
	12\qquad   \qquad \qquad $\textbf{end If}$\\
	13: \qquad   $\textbf{End For}$\\
	14:  \qquad Computing the error by \eqref{eq:error} \\
	15: \qquad \textbf{If} \qquad $\epsilon_n(\mathbf{C}_\mathrm{L})$ $\leq \delta$ \qquad \textbf{then}\\
	16: \qquad  Calculate IPNC matrix by \eqref{Final R_i+n} \\
	17: \qquad  Estimate the IPNC matrix, $\hat{\mathbf{R}}_{i+n} $ using the first $M$ row and $M$ column  of (17)   \\
	18: \qquad \textbf{End If}\\
	19: \textbf{End For} \\
	20: Design proposed beamformer using \eqref{proposed wegight vector}\\
	21: \textbf{Output:}\: Proposed beamforming weight vector $ \textbf{w}_{\text{prop}} $
\end{algorithm}
The computational complexity of the proposed LCSSP algorithm is $ \mathcal{O}(M^2L)$. The solution of the QCQP problem in \cite{gu2012robust} to obtain the optimal weight vector has complexity of at least $ \mathcal{O}(M^{3.5}) $,  while the beamformer in \cite{gu2019adaptive} has a complexity of $ \mathcal{O}(SM^{3}) $ and the reconstructed IPNC matrices in \cite{chen2015robust} and \cite{zhang2016interference} have a complexity of $ \mathcal{O}(M^3) $. Also, the cost of the beamformer in \cite{zheng2018covariance} is $ \mathcal{O}(\mathrm{max}(M^2S,M^{3.5})) $ and the beamformer in \cite{Saeed2020} needs $ \mathcal{O}(QM^2)$ complexity where $S$ is the uniform sampling points of the desired signal region.
\section{Simulations}
In this section, a uniform linear array with $ M=10 $ omnidirectional sensors is used. Three signals generated from complex white Gaussian noises are considered. It is assumed that there is one desired signal from the presumed direction  $\bar{\theta}_\mathrm{s}=0^\circ $ while the uncorrelated interference signals are impinging from  $-30^\circ$ and $30^\circ$.  The proposed LCSSP method is compared with the beamformer in \cite{chen2015robust} (IPNC-CC),  the reconstruction-estimation based beamformer in \cite{gu2012robust} (IPNC-Est), the beamformer in \cite{gu2019adaptive} (IPNC-OS), the beamformer in \cite{zhang2016interference} (IPNC-SPSS), the beamformer in \cite{zheng2018covariance} (IPNC-Re) and the beamformer in \cite{Saeed2020} (IPNC-MEPS). The input interference to noise ratios (INRs) of the two interferers are both set to 30 dB. In the IPNC-CC and IPNC-Est beamformers the number of sampling points for interference-plus-noise region is fixed at 200. In the beamformer IPNC-Re, the upper bound of the norm of the SV mismatch is set to $ \sqrt{0.1} $. In the proposed method, we employ $L = 20$, $K=50$ snapshots, and perform 100 Monte-Carlo runs. The angular sector of the desired signal is set to be $ {\Theta}=[{\bar{\theta}_\mathrm{s}}-6^\circ,{\bar{\theta}_\mathrm{s}}+6^\circ] $ where the interference angular sector is $ \bar{\Theta}=[-90^\circ,{\bar{\theta}_\mathrm{s}}-6^\circ)\cup({\bar{\theta}_\mathrm{s}}+6^\circ,90^\circ] $. 

In the first example, we compare the beampatterns of the reconstructed IPNC methods. We assume that the input SNR is fixed at 10 dB. Fig.~\ref{Beampattern} illustrates that all tested beamformers can steer the mainlobe to $\bar{\theta}_\mathrm{s}=0^{\circ}$. However, in the proposed LCSSP beamformer the desired signal is preserved and the interferers are effectively suppressed as the depth of the nulls is larger, which indicates that the interference suppression of LCSSP outperforms other methods.

In the second example, we evaluate the proposed method in the presence of the look direction mismatch and model mismatches due to the sensor displacement errors. In this example, the INR is fixed at 10 dB and it is assumed that the desired signal and the interferers are uniformly distributed in $ [-6^\circ,6^\circ] $ while the difference between the actual and assumed SV is modeled as array geometry errors, assuming the sensor position is drawn uniformly from $[-0.05,0.05]$ wavelength. Note that the DoA of the desired signal and the actual sensor position changes from run to run while remaining constant over samples.
\begin{figure}[t]
	\centering
	\includegraphics[height=2.4in]{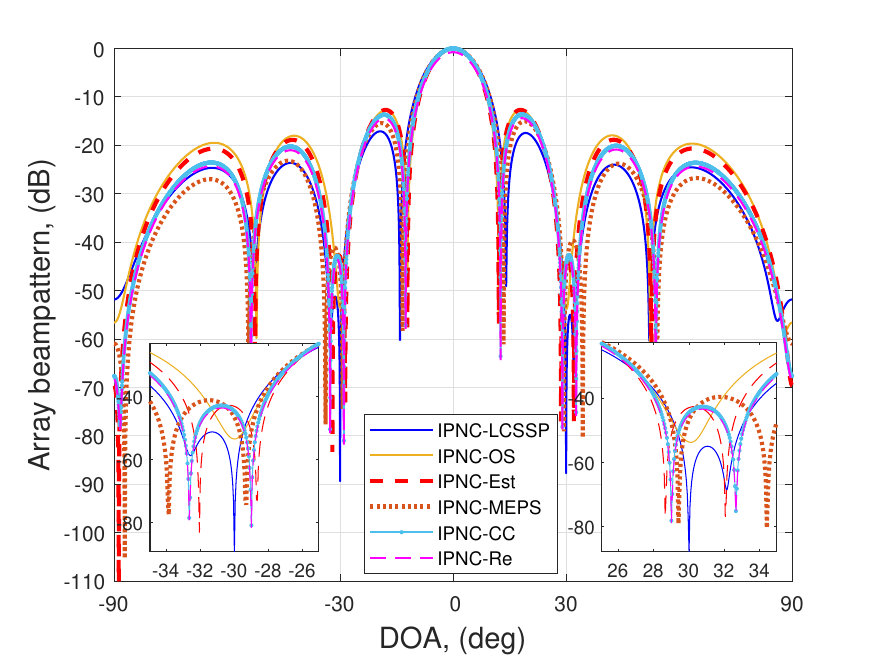}
\vspace{-0.75em}
	\caption{Comparison of the normalized beampatterns}
	\label{Beampattern}
\end{figure}
In Fig.~\ref{SNR}, we compare the SINR performance versus the SNR where the number of snapshots is fixed at $K=50$. Since the difference from the optimal SINR at low SNRs is not discernible, the deviations from the optimal SINR are displayed in  Fig.~\ref{Deviation}. From the results, it is observed that, because of random sensor position errors, there is an almost constant performance loss for IPNC-Est and IPNC-CC regardless of the input SNR. At the SNRs higher than 0 dB, the IPNC-Re beamformer has a performance loss because of the look direction mismatch. On the other hand,  the proposed LCSSP beamformer almost attains the optimal output SINR under these mismatches for all SNRs.
\begin{figure}[t]
	\centering
	\includegraphics[height=2.4in]{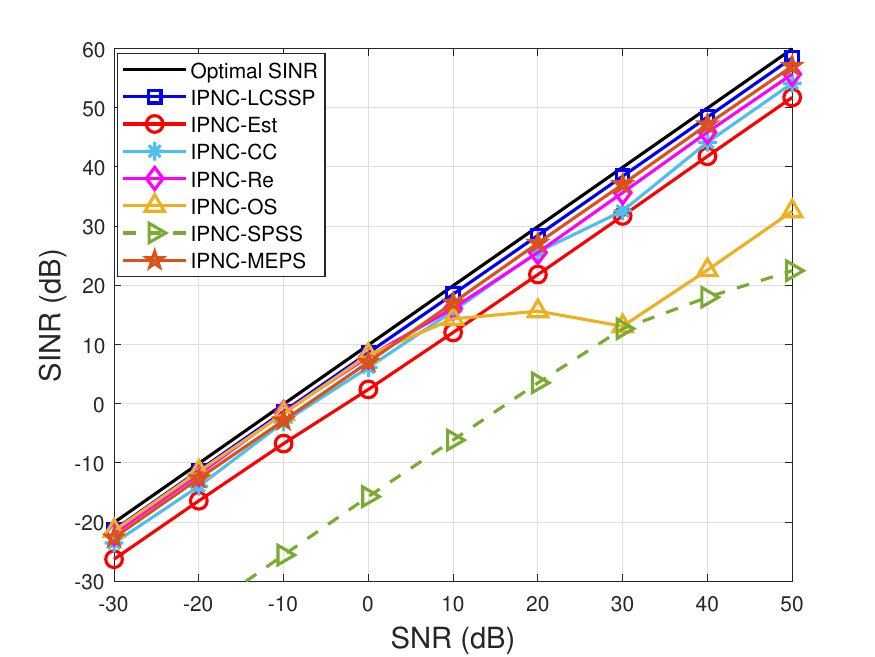}
\vspace{-0.75em}
	\caption{Output SINR versus Input SNR}
	\label{SNR}
\end{figure}
\begin{figure}[t]
	\centering
	\includegraphics[height=2.4in]{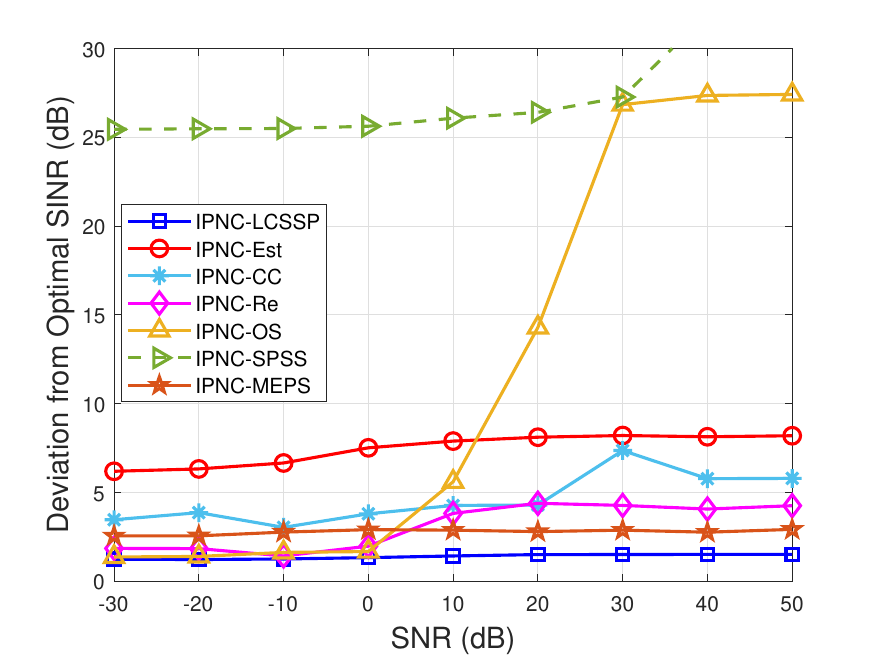}
\vspace{-0.75em}
	\caption{ Deviation from optimal SINR versus SNR }
	\label{Deviation}
\end{figure}
\begin{figure}[t]
	\centering
	\includegraphics[height=2.4in]{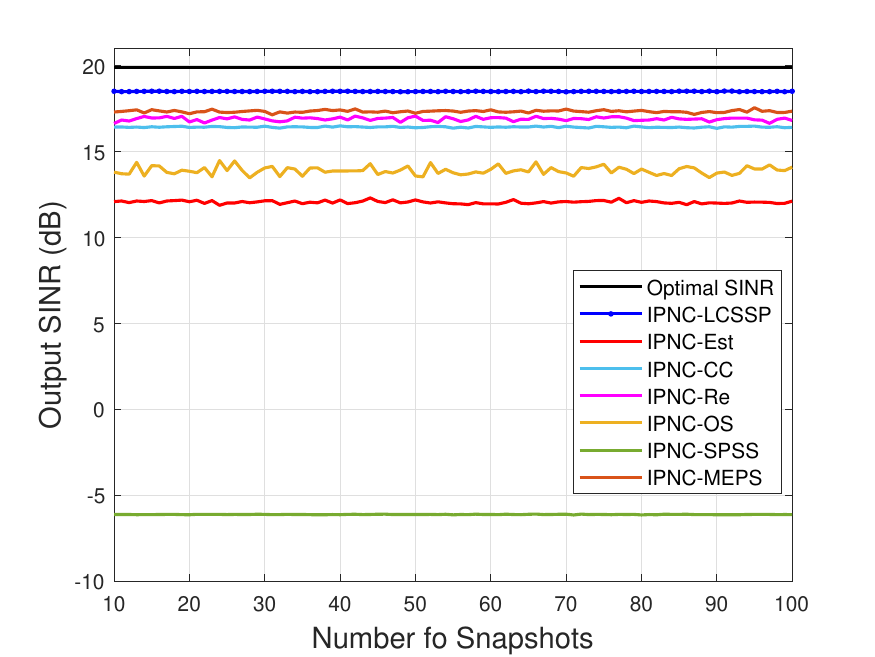}
\vspace{-0.75em}
	\caption{Output SINR versus Number of Snapshots}
	\label{Snapshots}
\end{figure}
\begin{figure}[!]
	\centering
	\includegraphics[height=2.4in]{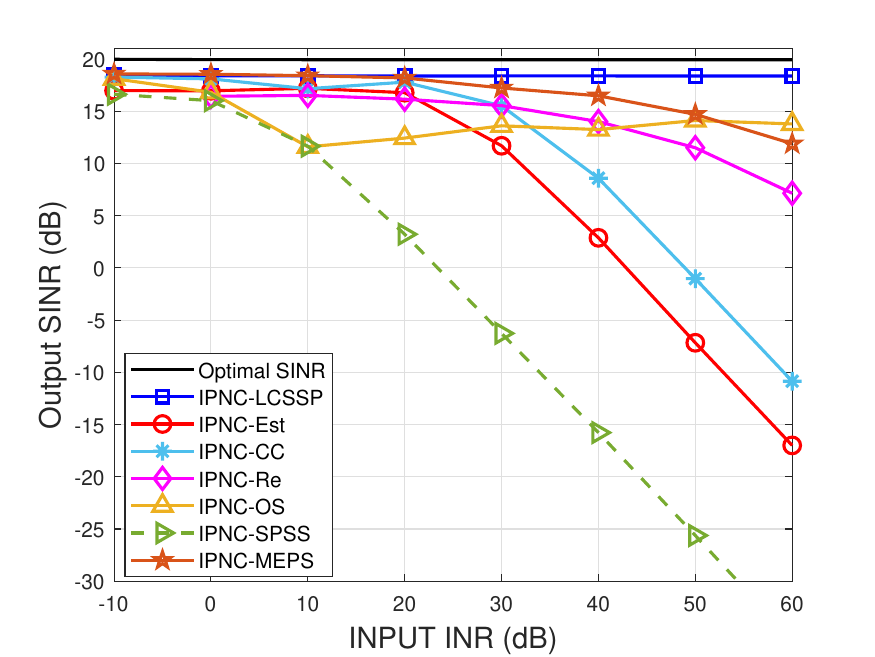}
\vspace{-0.75em}
	\caption{Output SINR versus input INR}
	\label{INR}
\end{figure}
In Fig.~\ref{Snapshots}, the performance of all tested beamformers is examined as the number of snapshots is increased. The excellent performance of LCSSP stems from its highly accurate estimate of the IPNC matrix without any signal power estimation, which enhances the robustness of IPNC-LCSSP against random look direction and array geometry errors over the snapshots.
In Fig.~\ref{INR}, we consider the case when the input INR varies, but the SNR is fixed at 10 dB. Clearly, the proposed LCSSP beamformer approaches the output SINR regardless of the interference power. However, both the IPNC-CC and IPNC-Est beamformers degrade as the interference power increases because of the array mismatch.

\section{Conclusion}
\label{sec:conclusion}
In this work, an efficient and accurate estimation of the IPNC matrix has been proposed using a low-complexity spatial sampling process and employing a virtual received array vector. In the proposed LCSSP approach, the power spectrum sampling has been carried out by a  projection matrix in a higher dimension. Moreover, the proposed LCSSP approach avoids estimation of the IPNC matrix by integrating over the angular sector of the interference. Simulation results have shown that the proposed LCSSP algorithm outperforms recently reported approaches.

\bibliographystyle{IEEEbib}
\bibliography{refs}

\end{document}